\documentclass[conference, compsocconf]{IEEEtran}
\usepackage[usenames,dvipsnames,svgnames,table]{xcolor}

\usepackage{tabularx}
\usepackage{pgfplots}
\usepackage{listings} 

\usepackage{url}
\usepackage[colorlinks=true,citecolor=Green]{hyperref}
\usepackage{lscape}
\usepackage[latin1]{inputenc}
\usepackage{tikz-timing}
\usepackage{tikz}
\usetikzlibrary{plotmarks}

\usepackage{caption}
\usepackage{subcaption}
\usepackage{sverb}
\usepackage[squaren,Gray]{SIunits}
\usepackage{amsmath} 
\usepackage{amsthm}
\usepackage{balance}

\makeatletter
\def\url@leostyle{%
  \@ifundefined{selectfont}{\def\UrlFont{\sf}}{\def\UrlFont{\small\ttfamily}}}
\makeatother
\lstdefinelanguage[ARM]{Assembler}%
  {morekeywords=[1]{.text,.globl,.align},%
  morekeywords=[2]{add, adr, sub, addc, adcs, adds, b, stmdb, %
    sub.W, beq,bge,blt,bne,%
    umull, umlal, cmp, ldr, mov, mul, pop, push, mrs, msr, %
    subs,vadd,vld1,vldm,vmov,vmul,%
    vpadd,vpop,vpush, str, b, bl,%
    itttt, ittee, itt, itte, it???, it??, itee, ittt, addne, subeq, moveq, subeq, movne, eorne, eor, movs,bx,
   },%
   morekeywords=[3]{f32,s32,i32},
   morekeywords=[4]{pc,r0,r1,r2,r3,r4,r5,r6,r7,r8,r9,r10,r11,r12,lr,sp,psp,control,basepri},
   keywordsprefix=.,%
   sensitive=false,%
   morecomment=[l]{;},
   moredelim=*[directive]\#,%
   moredirectives={define,elif,else,endif,error,if,ifdef,ifndef,line,%
      include,pragma,undef,warning}%
  }[keywords,comments,directives]

\definecolor{listing_background}{RGB}{250,250,250}
\definecolor{registers}{rgb}{0,0.4,0}
\definecolor{comments}{rgb}{0.4,0.4,0.4}

\renewcommand\IEEEkeywordsname{Keywords}

\makeatletter
\def\blfootnote{\xdef\@thefnmark{}\@footnotetext}
\makeatother

\theoremstyle{definition}
\newtheorem*{definition}{Definition}

\makeatletter

\newskip\@bigflushglue \@bigflushglue = -100pt plus 1fil

\def\bigcenter{\trivlist \bigcentering\item\relax}
\def\bigcentering{\let\\\@centercr\rightskip\@bigflushglue%
\leftskip\@bigflushglue
\parindent\z@\parfillskip\z@skip}

\makeatother

\title{Electromagnetic fault injection: towards a fault model on a 32-bit microcontroller}
\author{Blinded for review}

\makeatletter
\hypersetup{
    pdftitle = {\@title},
    pdfauthor = {Nicolas Moro, Amine Dehbaoui, Karine Heydemann, Bruno Robisson, Emmanuelle Encrenaz},
    pdfsubject = {FDTC 2013},
    pdfkeywords = {microcontroller, timing fault, electromagnetic glitch, fault attack, fault model}
}

\begin{document}
\makeatother

\author{
    \IEEEauthorblockN{Nicolas Moro\IEEEauthorrefmark{1}\IEEEauthorrefmark{3}, Amine Dehbaoui\IEEEauthorrefmark{2}, Karine Heydemann\IEEEauthorrefmark{3}, Bruno Robisson\IEEEauthorrefmark{1}, Emmanuelle Encrenaz\IEEEauthorrefmark{3}}

    \IEEEauthorblockA{\IEEEauthorrefmark{1}Commissariat à l'Énergie Atomique et aux Énergies Alternatives (CEA)
    \\Gardanne, France
    \\Email: \{nicolas.moro, bruno.robisson\}@cea.fr}
    \IEEEauthorblockA{\IEEEauthorrefmark{2}École Nationale Supérieure des Mines de Saint-Étienne (ENSM.SE)
    \\Gardanne, France
    \\Email: amine.dehbaoui@mines-stetienne.fr}
    \IEEEauthorblockA{\IEEEauthorrefmark{3}Laboratoire d'Informatique de Paris 6 (LIP6)
    \\Sorbonne Universités, UPMC Univ Paris 06, UMR 7606, LIP6
    \\Paris, France
    \\Email: \{nicolas.moro, karine.heydemann, emmanuelle.encrenaz\}@lip6.fr
	}
}

\maketitle

\begin{abstract}
Injection of transient faults as a way to attack cryptographic implementations has been largely studied in the last decade. Several attacks that use electromagnetic fault injection against hardware or software architectures have already been presented. On microcontrollers, electromagnetic fault injection has mostly been seen as a way to skip assembly instructions or subroutine calls. However, to the best of our knowledge, no precise study about the impact of an electromagnetic glitch fault injection on a microcontroller has been proposed yet. The aim of this paper is twofold: providing a more in-depth study of the effects of electromagnetic glitch fault injection on a state-of-the-art microcontroller and building an associated register-transfer level fault model.
\end{abstract}


\blfootnote{© 2013 IEEE. Personal use of this material is permitted. Permission from IEEE must be obtained for all other uses, in any current or future media, including reprinting/republishing this material for advertising or promotional purposes, creating new collective works, for resale or redistribution to servers or lists, or reuse of any copyrighted component of this work in other works.}

\section{Introduction}

Physical attacks aim at breaking cryptosystems by gaining information from their implementation instead of using theoretical weaknesses. Those attack schemes were introduced in the late 1990s. There are two main subclasses of physical attacks: passive and active ones. In passive attacks, an attacker uses the fact that some measurable data may leak information about manipulated secret data such as cryptographic keys. Physical quantities which can be used for passive attacks include execution time \cite{Kocher1996}, electromagnetic radiations \cite{Agrawal2003}, power consumption \cite{Kocher1999} or light emissions \cite{Schlosser2012}. In active attacks, an attacker modifies the circuit's behaviour in order to perform its attack scheme. Fault attacks are a subset of active attacks in which an attacker injects a transient fault in a circuit's computation.

Faults attacks were introduced in 1997 by Boneh \textit{et al.} \cite{Boneh1997}. They consist in modifying a circuit environment in order to change its behaviour or to induce faults into its computations \cite{BarEl2006} \cite{Karaklajic2009} \cite{Barenghi2012}. Many means are of common use to inject such faults, especially laser shots \cite{Skorobogatov2003} \cite{Schmidt2007} \cite{Canivet2010}, overclocking \cite{Agoyan2010b} \cite{Balasch2011}, chip underpowering \cite{Fournier2003} \cite{Zussa2012}, temperature increase \cite{Skorobogatov2009} or electromagnetic glitches \cite{Schmidt2007} \cite{Dehbaoui2012}. 

There are three main subclasses of fault attacks: algorithm modifications, safe error and differential fault analysis. Algorithm modifications aim at skipping \cite{Schmidt2008} or replacing \cite{Balasch2011} some instructions executed by a microcontroller to circumvent its security features. Safe-error attacks aim at evaluating whether or not a fault injection has an impact on the output \cite{Yen2000}. Differential fault analysis (\textsc{DFA}) aims at retrieving the keys used by an encryption algorithm by comparing correct ciphertext and faulty ciphertexts (i.e. ciphertexts obtained from a faulted encryption). This technique was first introduced for public key encryption algorithms \cite{Boneh1997}, and quicky extended to secret key algorithms \cite{Biham1997}. 

From that time, many attack schemes have been proposed to attack various encryption algorithms. They all rely on an attacker's fault model which defines the type of faults the attacker can perform \cite{Barenghi2010}. Thus, they require a high accuracy in the fault injection process. If the faults are not induced at the proper time in the algorithm, or affect the wrong bits, the entire attack process fails. As a consequence, the ability to precisely control the fault injection process is a key element in carrying out any fault attack. Common fault models include instruction skips \cite{Schmidt2008}, single bit faults or single word faults \cite{Verbauwhede2011}.


In this work, we report the use of electromagnetic pulses to induce faults into the computations of an up-to-date microcontroller. We also report a study of the local effect of electromagnetic pulses. Moreover, the underlying effects behind common fault models are not always clearly understood and may highly depend on the target architecture. As a consequence, this work finally aims at defining a precise fault model and providing an understanding of the faults an electromagnetic glitch can induce on an embedded program.

\smallskip
The rest of this paper is organized as follows. Section \ref{Section:Approche} introduces our fault injection experimental setup and details the approach we use. Section \ref{Section:Resultats} describes the influence of some experimental parameters on injected faults. Section \ref{Section:Fautes-obtenues} details the effects of the injected faults on the program flow and data flow. Finally, the resulting register-transfer level fault model is presented in section \ref{Section:Modele}. Section \ref{Section:Related works} gives details about some related research papers.

\section{Approach}
\label{Section:Approche}

This section starts by describing our experimental setup choices in \ref{Paragraphe:ExperimentalSetup}. This experimental setup enables us to provide the results presented in section \ref{Section:Resultats} which show the influence of the different experimental parameters. Then, we detail the approach we use to precisely characterize the faults we injected. This characterization method, which matches experimental results obtained from the microcontroller with simulation data, is detailed in \ref{Paragraphe:Procede}.

\subsection{Experimental setup}
\label{Paragraphe:ExperimentalSetup}

\subsubsection{Electromagnetic fault injection bench}
\begin{figure}[!h]
\centering
\includegraphics[scale=0.7]{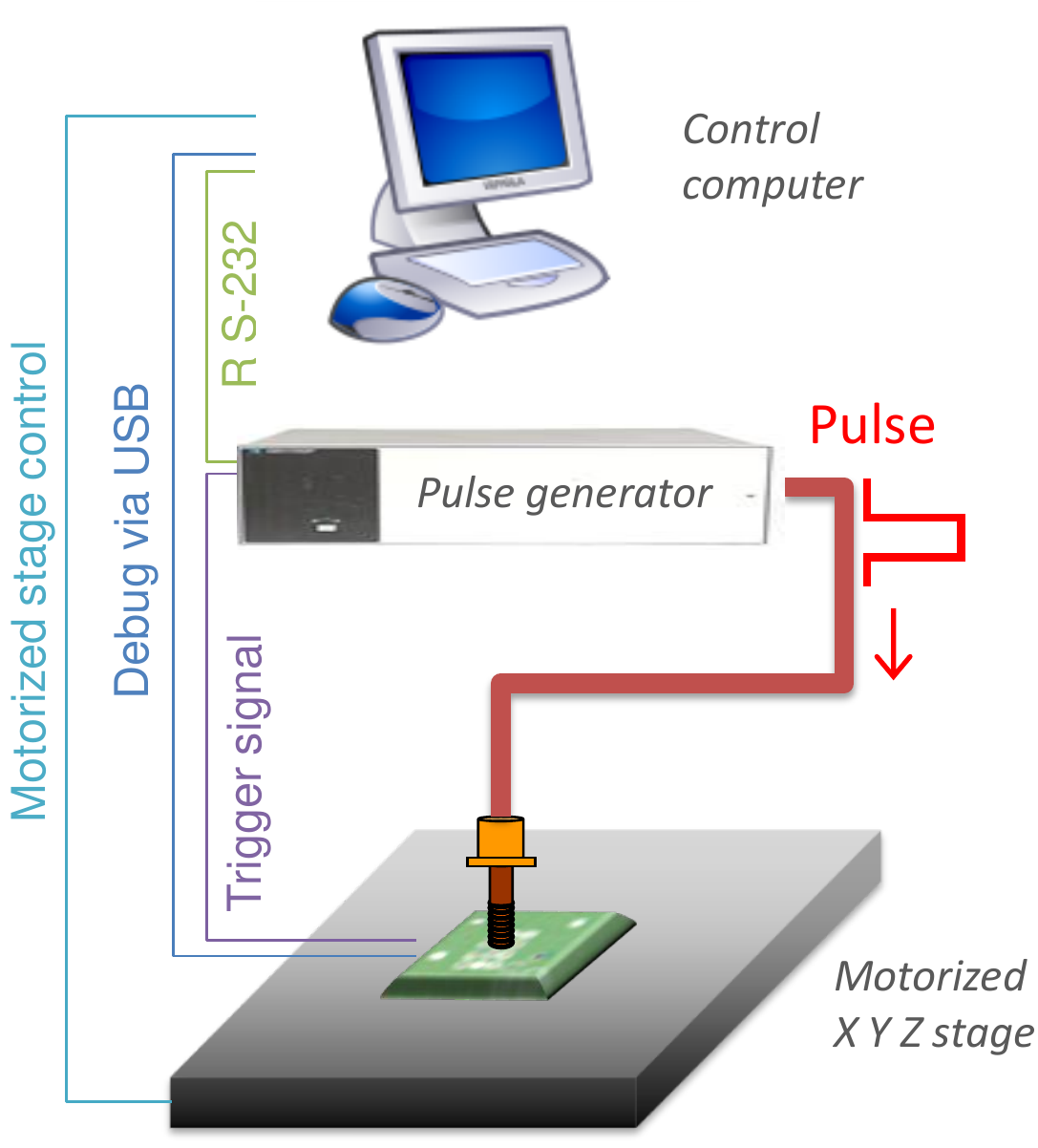}
\caption{Electromagnetic fault injection bench}
\label{Image:Banc}
\end{figure}

The electromagnetic glitch fault injection platform shown in Fig. \ref{Image:Banc} is composed of a control computer, the target device, a motorized stage, a pulse generator, and a magnetic antenna. The target (described in \ref{Paragraphe:Target}) is mounted on the X Y Z motorized stage. The computer controls both the pulse generator (through a RS-232 link) and the target board (through a USB link).

The pulse generator is used to deliver voltage pulses to the magnetic coil. It has a constant rise and fall transition time of \unit{2}{\nano\second}. The amplitude range of the generated pulses extends from \unit{-200}{\volt} to \unit{200}{\volt}, their width extends from \unit{10}{\nano\second} to \unit{200}{\nano\second}. The magnetic antenna we use is composed of a few turns with a diameter of \unit{1}{\milli\meter}. We use it in order to disturb a small part of the target device. This spatial accuracy is possible thanks to a high accuracy X Y Z stage.

\subsubsection{Target}
\label{Paragraphe:Target}
The chosen target is an up-to-date 32-bit microcontroller designed in a CMOS \unit{130}{\nano\meter} technology. It is based on the ARM Cortex-M3 processor \cite{DefinitiveGuideARMCortexM3}. Its operating frequency is set to \unit{56}{\mega\hertz}. This microcontroller does not embed any cache memory.

\paragraph{Choice of target}
The target we use is a state-of-the-art microchip, based on a recent technology. ARM Cortex processors are already very widespread for both mainstream and secure microcontrollers. Although we did not choose a smartcard version of the microcontroller, this target embeds some basic security mechanisms against clock perturbations and voltage glitches. Moreover, several interrupt vectors have been defined which can handle some hardware faults and can be used for a basic fault detection. Hence, we can consider this target as reasonably secured against some of the most common low-cost fault injection means. However, this target does not embed any protective shield against reverse engineering or electromagnetic injection. Since this research aims at understanding the effects of fault injection on a recent microcontroller, we do not work on a highly-secure version of this microcontroller.

\paragraph{Instruction set}
Cortex-M3 processors run the ARM Thumb2 instruction set \cite{Thumb2}. Thumb2 is actually the successor to both ARM and Thumb instruction sets, and contains both 16-bit and 32-bit instructions. 

\paragraph{Hardware interrupts}
\label{Paragraphe:Interruptions}
Several fault exceptions can catch illegal memory accesses or illegal program behaviour. Those fault exceptions are \texttt{Hard Fault}, \texttt{Bus Fault}, \texttt{Usage Fault} and \texttt{Memory Management Fault}. Each of these exceptions can be triggered for several subtypes of hardware faults. In the following experiments, every exception handler function executes an infinite loop.

\subsection{Experimental process}
\label{Paragraphe:Procede}

Working with a microcontroller in such a black-box approach requires to develop a specific experimental approach. This approach aims at enabling us to deduce the effects of faults by observing some internal data from the microcontroller. This observation must be done with a non-invasive technique. Since a faults may have an impact on the program flow and since we need to access some accurate data such as registers or cycle count, the communication cannot be done with a serial link. We use the JTAG-equivalent non-instrusive SWD debug link to retrieve data from the microcontroller. Besides, we also use the hardware exceptions defined in \ref{Paragraphe:Interruptions} as a way to get some extra data about the injected faults.

\subsubsection{Microcontroller's internal state observation}

The experimental measurement process we use is the following:
\begin{itemize}
\item Reset the microcontroller
\item Execute the target code
\item Send a pulse to the injection antenna
\item Interrupt the program execution
\item Harvest the microcontroller's internal data 
\end{itemize}

The following paragraphs detail the important elements of this experimental process.

\paragraph{Trigger window}
In order to have a correct view of the microcontroller's internal data, we have created an assembly subroutine containing some test instructions (which will be detailed in section \ref{Section:Resultats}). For our experiments, the microcontroller sets a trigger signal for the electromagnetic injection. With this technique, we can target the executed program at the scale of a single instruction. By observing the microcontroller's clock during this trigger window, we can focus the injection on a single clock cycle. Besides, the pulse injection time is defined by reference to the beginning of the trigger signal temporal window.

\paragraph{Watchpoint and program end}
In this experimental process, the program normally stops because of a breakpoint set after the target code. This watchpoint is defined before popping the stack at the end of the assembly subroutine and after the trigger window. However, with our experimental setup and target code, two other scenarios may happen because of a fault: an unconditional jump and an infinite loop due to the triggering of an exception. These two scenarios modify the control flow, and the program may not reach the defined breakpoint. Moreover, the unconditional jump scenario makes the setting of breakpoints very hard. To handle these issues, our control computer stops the microcontroller after a fixed delay. 

\paragraph{Internal data}
With the SWD debug link, the internal data we get from the microcontroller at a watchpoint for our experiments are: the general-purpose registers (\texttt{r0} to \texttt{r12}), the stack pointer (\texttt{r13}), the link register (\texttt{r14}), the program counter (\texttt{r15}), the program status register (\texttt{xPSR}), some chosen variables in memory and the number of clock cycles taken by our experiment. This number of clock cycles is counted from the beginning of the target subroutine. The \texttt{xPSR} register gives us information about the processor flags and the exceptions that may have been triggered. Since we only inject transient faults and since the watchpoint is set several clock cycles after our attack, we can reasonably assume that the debugging module embedded in the chip is not corrupted when recovering the internal data from the microcontroller. 

\smallskip
However, some internal data such as the instruction register cannot be accessed. When working at the scale of a single instruction, we may need to determine which instruction has been actually executed by the core. To get a list of suitable instructions, we need to rely on an exhaustive instruction simulation.

\subsubsection{Fault model simulation}
\label{Paragraphe:Simulation}

\begin{figure}[!h]
\includegraphics[scale=0.45]{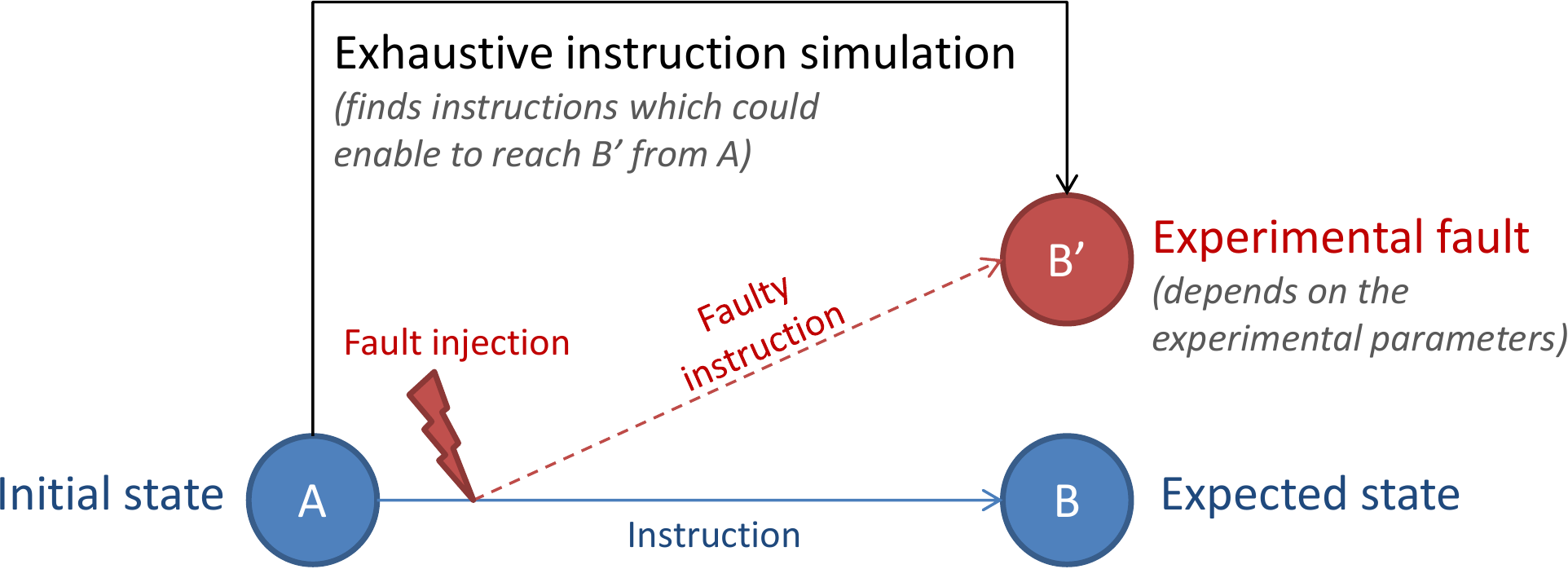}
\caption{Our approach to characterize the injected faults}
\label{Image:Approche}
\end{figure}

We propose to use simulation to explain the effects of electromagnetic fault injection. Our approach aims at comparing the experimental faulty outputs with outputs from a fault model simulation. Thus, we can validate the interpretation of these effects by comparing the outputs with the internal data. This scheme is summarized in Fig. \ref{Image:Approche}. 

Simulations aim at finding output states which could be compatible with the output states we observed. In order to match simulations with measurements, we define a binary relation between experimental output states and simulated output states. 

\begin{definition}
One instruction replacement \textbf{can explain} an experimental measurement if the output states ([\texttt{r0}-\texttt{r12}], \texttt{xPSR}) at the defined watchpoint are the same for the measurement and the simulation
\end{definition}

In the rest of this article, two classes of faults can be distinguished: faults on the \textit{data flow} and faults on the \textit{program flow} \cite{Verbauwhede2011}. Faults that lead to the replacement of an instruction by another one are faults on the program flow. They may result in an algorithm modification, depending on the context and the replaced instruction. On the contrary, faults which only modify a piece of data without modifying an instruction are faults on the data flow. 

Nevertheless, this difference might not be clear for many cases since both fault classes may lead to very similar visible outputs. Thus, defining whether a resulting faulty output is a consequence of a fault on the data flow or the control flow is generally a tough task. Nevertheless, a single assembly instruction can only output a very limited set of data. As a consequence, it is possible to tell whether or not a faulty output is the consequence of a fault on the control flow. Thus, every faulty output which cannot be explained by an instruction replacement is considered to come from a fault on the data flow.


The Thumb2 instruction set is composed of both 16-bit and 32-bit instructions. 16-bit instructions can be exhaustively tested. 32-bit instruction start with the prefixes \texttt{11101} or \texttt{1111}, which reduces the complexity of an exhaustive test. Moreover, the 32-bit part of the instruction set is mostly sparse, we can remove many branches in the search space. 

\begin{figure*}[!t]
\bigcenter
\includegraphics[scale=0.7]{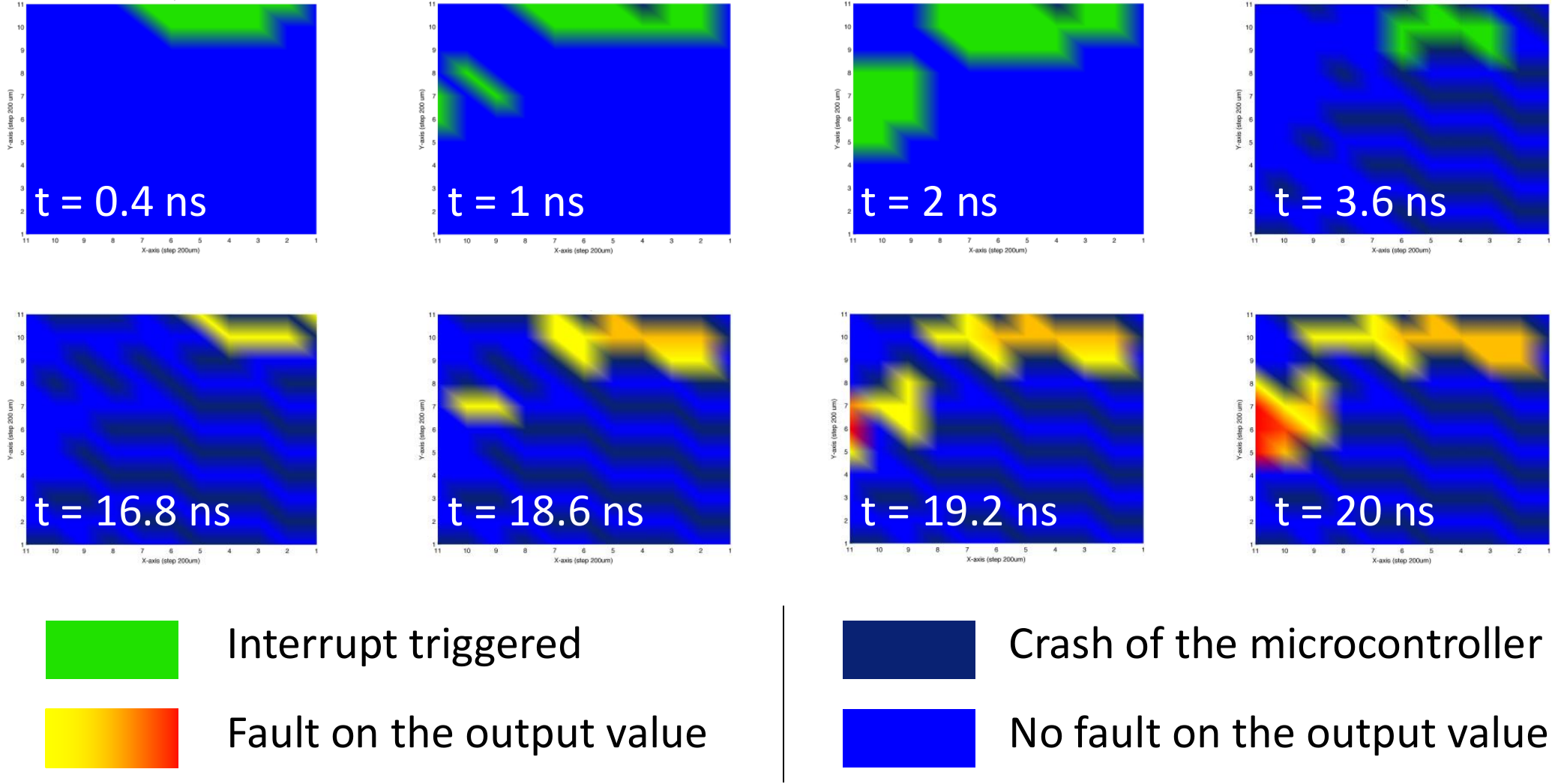}
\caption{Impact of the probe's position}
\label{Image:Carto-Spatiale}
\end{figure*}

It should be noted that this simulation is performed on the same binary as the one that is used for the fault injection experiments. To perform this simulation, we developed a specific program, based on the Keil UVSOCK library. Our simulation program is able to control the Keil \micro{}Vision debugger during an execution on the Keil \micro{}Vision simulator. It emulates faults on the control flow by replacing on the fly the target instruction. 

Obviously, many instruction replacements may be able to explain one single measurement. Nevertheless, being able to simulate instruction replacement will enable us to explain the effects we observed and then to define a fault model more clearly. To sum up, an exhaustive simulation over the instruction set is practical and can be performed in real conditions. Moreover, it enables to distinguish faults on the control flow from faults on the data flow.

\section{Experimental study of the injection parameters}
\label{Section:Resultats}

In this section, we provide a study of the influence of several experimental parameters on the final outputs. Since metastability phenomena appear, we first start by describing them in the following paragraph.

\subsection{Metastability phenomena}
\label{Paragraphe:Metastabilite}
Since electromagnetic glitch fault injection leads to timing faults \cite{Dehbaoui2012}, we obtained some metastability phenomena. For this experiment, the pulse's voltage was set to \unit{190}{\volt}, the clock frequency was set to \unit{56}{\mega\hertz}, the pulse's injection time was fixed to an arbitrary value, and the pulse width was set to \unit{10}{\nano\second}. The probe position was found by a trial-and-reset approach. The results for 10000 executions of our experimental process are presented in Table \ref{Tableau:Metastabilite}, every observed output value is associated to its occurrence rate. They show a metastability phenomenon for a single load instruction from the Flash memory which correct loaded value is \texttt{0x12345678} since several values appear for the same fixed configuration of the experimental parameters.

\begin{table}[!h]
\centering
\caption{Metastability phenomenon for a single load instruction}

\begin{tabular}{|l|l|} 
   \hline
   \textbf{Loaded value} & \textbf{Occurrence rate} \\
   \hline
   1234 5678 (no fault) & 60.1\% \\
   \hline
   \textcolor{red}{FFF}4 567\textcolor{red}{9} & 27.4\% \\
   \hline
   \textcolor{red}{FFFC} 567\textcolor{red}{9} & 12.3\% \\
   \hline   
   \textcolor{red}{FFFC} 567\textcolor{red}{b} & 0.1\% \\
   \hline   
   \textcolor{red}{FFFC 7}67\textcolor{red}{9} & 0.1\% \\
   \hline   
\end{tabular}
\label{Tableau:Metastabilite}
\end{table}

\subsection{Study of the injection parameters}

In the case of an electromagnetic fault injection on a microcontroller, many experimental parameters can have an influence on the final outputs. The main parameters we can control in these experiments and which may have an influence are detailed in Table \ref{Tableau:ParametresExperimentaux}. For all the following experiments except the one that studies the voltage's influence, the pulse voltage was set to \unit{190}{\volt}. The pulse width was set to \unit{10}{\nano\second}, which is shorter than the \unit{17}{\nano\second} clock period (for a \unit{56}{\mega\hertz} clock frequency). In the following paragraphs, we detail the separate influence of some of these parameters.

\begin{table}[!h]
\centering
\caption{Experimental parameters}
\begin{tabular}{|l|l|} 
   \hline
   \textbf{Electromagnetic} 
   & - x-y-z position of the injection probe\\
   \textbf{injection parameters}   
   & - Pulse injection time \\
   & - Pulse characteristics (width, voltage) \\ 
   \hline
   \textbf{Microcontroller} 
   & - Operating frequency \\
   \textbf{hardware parameters}   
   & - Power supply\\
   \hline
   \textbf{Microcontroller} 
   & - Type of the executed instructions\\
   \textbf{software parameters}   
   & - Program memory (RAM or Flash) \\
   \hline
\end{tabular}
\label{Tableau:ParametresExperimentaux}
\end{table}

\subsubsection{Position of the injection probe over the package's surface}
\label{Paragraphe:Carto-spatiale}

\begin{figure*}[!t]
\bigcenter
\includegraphics[scale=0.7]{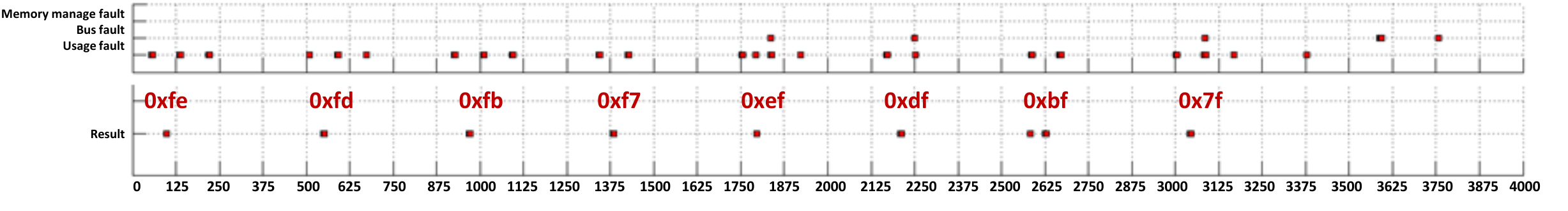}
\caption{Influence of the pulse's injection time for an array sum whose expected result is 0xFF}
\label{Image:Carto-AdditionTableau}
\end{figure*}

The X Y Z stage we use for our experiments enables us to vary the injection probe's position. Since varying the Z position of the antenna leads to a similar class of effects on the microcontroller than varying the pulse's voltage \cite{Carlier2012}, we fix a position for Z and only study the influence of the X Y position. In this experiment, we change the X Y coordinates and the pulse's injection time. This experiment is performed at the scale of a single load instruction which loads the value \texttt{0x12345678} from the Flash memory into the register \texttt{R8}. This fault injection has been performed over a \unit{20}{\nano\second} time interval, by steps of \unit{200}{\pico\second}. The probe browsed a \unit{3}{\milli\meter} square over the circuit's die, by steps of \unit{200}{\micro\meter}. Fig. \ref{Image:Carto-Spatiale} shows the results for this experiment. 

The experiment shows that there are four kinds of outputs, depending on the probe position and the injection time : no fault on the loaded value, a crash of the microcontroller, the triggering of a \texttt{Usage Fault} exception, and a fault on the value in \texttt{R8}. Very few faults on the register \texttt{R0} were also observed. Except from \texttt{R8} and \texttt{R0}, no other register was faulted in this experiment. Moreover, those two registers were never faulted together. Every faulty output we observed on \texttt{R8} has a higher Hamming weight than the \texttt{0x12345678} expected value. On Fig. \ref{Image:Carto-Spatiale}, yellow areas led to a small increase in this Hamming weight and red areas led to a high increase. This experiment highlights the local effect of electromagnetic fault injection on a microcontroller, with different effects depending on the probe's position. Since very few probe positions can lead to a successful fault injection, this spatial cartography also helps us to find some suitable X Y Z configurations for the following experiments.

\subsubsection{Injection time}
This experiment has been performed on the following test program for a fixed X Y Z position. This program uses a loop to sum the elements of an array that contains eight powers of two. \texttt{array[i]} contains $2^i$. At the end of the computation, the result stored at the address pointed by \texttt{r0} contains $0$x$FF$. This test program requires about \unit{3.5}{\micro\second} to complete. We performed this fault injection over this time interval, by steps of  \unit{200}{\pico\second}.

\begin{lstlisting}[language={[ARM]Assembler}]
addition_loop:
ldr	r4, [r2,r1, lsl #2] ; r4 = array[i]
ldr	r3, [r0,#0]	 				; r3 = result
add	r3, r4		 					; r3 = r3 + r4  
str	r3, [r0,#0]	 				; result = r3
add	r1, r1, #1	 				; r1 = r1 + 1
cmp	r1, #8		 					; r1 == 8 ?
blt	addition_loop
\end{lstlisting}

This test program enables us to perform an electromagnetic fault injection on a sample made of different instructions. The results for this experiment are shown in Fig. \ref{Image:Carto-AdditionTableau}. Three kinds of situations have been observed:
\begin{itemize}
\item \texttt{BusFault} or \texttt{UsageFault} hardware interrupts
\item A fault on the output value
\item A normal behaviour with no fault
\end{itemize}

Every fault we observed on the output value corresponds to an execution in which only one power of two has not been added. However, many faults  could explain such results. That is why the precise effect of electromagnetic fault injection at the scale of a single instruction is studied precisely in section \ref{Section:Fautes-obtenues}.

\subsubsection{Pulse characteristics}
In the following paragraphs, we study the separate influence of the pulse parameters. For these paragraphs, a fixed position was set for the injection probe. This position had been found thanks to the spatial cartography presented in \ref{Paragraphe:Carto-spatiale}. 


\paragraph{Pulse width}
The pulse width does have an influence on the outputs. According to Faraday's law of induction, the electromotive force induced in a loop (e.g. inside the power grid) corresponds to the time-derivative of the magnetic flux transmitted by the injection antenna. This magnetic flux is proportional to the current sent into the injection solenoid. Thus, the electromagnetic glitch that is transmitted to the circuit depends on the current's variations. We also observed that sending longer pulses reduces the stress applied to the circuit.

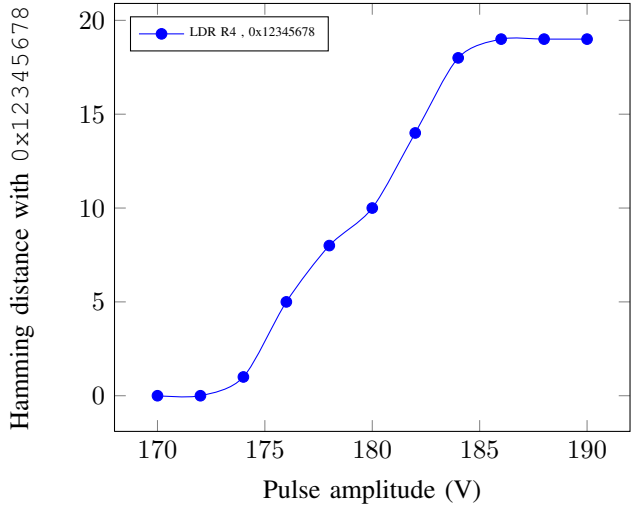
\begin{figure}[t!]

\begin{tikzpicture}
    \begin{axis}[
        xlabel=Pulse amplitude (V),
        ylabel=Hamming distance with \texttt{0x12345678},/pgfplots/enlargelimits=true,legend style={legend pos=north west,font=\tiny}]
    \addplot[smooth,mark=*,blue] plot coordinates {
(170,0)
(172,	0)
(174,	1)
(176,5)
(178,8)
(180,	10)
(182,	14)
(184,	18)
(186,19)
(188,19)
(190,19)};
    \addlegendentry{LDR  R4 , 0x12345678}
    \end{axis}
    \end{tikzpicture}
    \caption{Hamming distance with \texttt{0x12345678} versus pulse's voltage}
\label{Misc:Amplitude-HD}
    \end{figure}

\paragraph{Pulse voltage}
\label{Paragraphe:Impact-amplitude}
To evaluate the influence of the pulse voltage, the test program has been set to a single \texttt{LDR} assembly instruction. \texttt{LDR R\_{}o,[R\_{}i,\#{}offset]} loads the value pointed by \texttt{R\_{}i} with offset \texttt{\#{}offset} into the register \texttt{R\_{}o}. For the test instruction, the register \texttt{R\_{}i} pointed to a Flash memory address. To perform an analysis of the impact of the pulse's voltage, we needed to fix a suitable configuration for the other parameters. Those other parameters were set to some fixed values: we chose a configuration in which a fault occurs on the loaded value. For this experiment, the tested instruction was \texttt{LDR R4,[PC,\#{}44]}. The inital value of \texttt{R4} was \texttt{0x0} and \texttt{PC$+$44} was a Flash memory address which contained \texttt{0x12345678}. Since metastability phenomena appear, for this experiment we take into account the faulty output with the highest occurrence rate. Table \ref{Tableau:ImpactAmplitude} shows the value in \texttt{R4} for different values of the pulse voltage. According to those results, increasing the pulse voltage increases the Hamming weight of the loaded value. This pattern is highlighted by Fig. \ref{Misc:Amplitude-HD}, which shows the Hamming distance with the \texttt{0x12345678} expected value versus the pulse's voltage. The same kind of trend has been obtained for different values for the probe position and the injection time. However, it seems that only instructions which loads a value from the Flash memory can lead to this kind of \textit{set at 1} fault. Indeed, we did not manage to inject similar faults in case of a data transfer from the SRAM memory.

\begin{table}[!h]
\centering
\caption{Influence of the pulse's voltage}
\begin{tabular}{|l|l|l|} 
   \hline
   \textbf{Pulse voltage} & \textbf{Loaded value}  & \textbf{Occurrence rate} \\
   \hline
   \unit{170}{\volt} & 1234 5678 (no fault) & 100\% \\
   \hline
   \unit{172}{\volt} & 1234 5678 (no fault) & 100\%\\
   \hline
   \unit{174}{\volt} & \textcolor{red}{9}234 5678 & 73\% \\
   \hline   
   \unit{176}{\volt} & \textcolor{red}{FE}34 5678 & 30\% \\
   \hline   
   \unit{178}{\volt} & \textcolor{red}{FFF}4 5678 & 53\% \\
   \hline   
   \unit{180}{\volt} & \textcolor{red}{FFFD} 5678 & 50\% \\
   \hline   
   \unit{182}{\volt} & \textcolor{red}{FFFF 7F}78 & 46\% \\
   \hline   
   \unit{184}{\volt} & \textcolor{red}{FFFF FFFB} & 40\% \\
   \hline   
   \unit{186}{\volt} & \textcolor{red}{FFFF FFFF} & 100\% \\
   \hline
   \unit{188}{\volt} & \textcolor{red}{FFFF FFFF} & 100\% \\
   \hline
   \unit{190}{\volt} & \textcolor{red}{FFFF FFFF} & 100\% \\
   \hline
\end{tabular}
\label{Tableau:ImpactAmplitude}
\end{table}

\subsubsection{Type of the executed instructions}
Our experiments highlighted a significant trend: we managed to inject faults on different types of instructions such as branch instructions, ALU instructions or load-store instructions. However, load instructions from the Flash memory were significantly easier to fault. The microcontroller we use has a Harvard architecture. Every instruction fetch uses the instruction bus. Moreover, load instructions also use the data bus in the decode pipeline phase. As a consequence, section \ref{Section:Fautes-obtenues} provides a more detailed study of the consequences of this fault injection in two cases: one case to highlight a fault on the instruction bus, another one to highlight a fault on the data bus. On the one hand, we study the effects on a generic single instruction. On the other hand, we study the specific case of a load instruction from the Flash memory.

\section{Experiments on the data and instruction bus}
\label{Paragraphe:Fautes-obtenues}
\label{Section:Fautes-obtenues}

The two following subsections detail the results we obtained when trying to inject faults into the control flow or the data flow of the target program. In order to minimize the side effects which may happen when studying a big number of assembly instructions, the following results have been obtained for two classes of test applications. To highlight faults on the control flow, we use a sequence of \texttt{NOP} instructions \cite{Spruyt2012}. Since \texttt{NOP} instructions have no effect, a faulty output will be easier to notice and to explain. To highlight faults on the data flow, the test application we use is a single \texttt{LDR} instruction which loads data from memory into a register. The initial values at the beginning of our target function are detailed in Table \ref{Tableau:ValeursInitiales}. Those beginning values are the same for the two following experiments. The comparison between the initial values, the output ones and the expected ones helps us to have a better understanding of possible instruction replacements effects.  
 
\begin{table}[!h]
\centering
\caption{Initial values at the beginning of the execution}
\begin{tabular}{|l|l|} 
   \hline
   \textbf{Piece of data} & \textbf{Value} \\
   \hline
   \texttt{r0} & A memory address in RAM \\
   \hline
   \texttt{r1} to \texttt{r4} & 0x1 to 0x4 \\
   \hline
   \texttt{r5} and \texttt{r6} & Not relevant \\
   \hline
   \texttt{r7} & 0x100 \\
   \hline   
   \texttt{r8} to \texttt{r12} & 0x00 \\
   \hline
   Address pointed by \texttt{r0} & 0x00 \\
   \hline   
   
\end{tabular}
\label{Tableau:ValeursInitiales}
\end{table}

\subsection{Faults on the program flow}
\label{Paragraphe:Fautes-ProgramFlow}

Faults on the program flow can be observed through instruction replacement faults thanks to the simulation. However, studying instruction replacement with two possible instruction sizes is a very tough task. Since every fetch from the code memory is 32-bit wide, we need to consider several instruction replacement scenarios. With this approach, we can simulate the replacement of a 16-bit or 32-bit instruction by another 16-bit or 32-bit instruction. However, two 16-bit instructions might be replaced by two different 16-bit instructions. Similarly, a 32-bit instruction might be replaced by two 16-bit instructions. Those two cases would imply performing an almost-exhaustive search over 32 bits, which is not practical in our case. Though, we could partially bypass this problem by recording the number of clock cycles in our experiments. However, guessing the number of executed instructions from the clock cycle count is not an easy task because of the complex instruction set. Observing this clock cycle count could theoretically enable us to exclude some replacement scenarios in further experiments. To highlight the possibility to inject faults on the program flow, the following experiment targeted a \texttt{NOP} sled. Since different position probes and different injection times lead to different results, the following results have been found for different experimental configurations of these parameters.

\subsubsection{Hardware exceptions}
Our fault injection sometimes led to an exception triggering. However, only \texttt{Usage Fault} exceptions were observed. More precisely, the \texttt{No coprocessor} exception and the \texttt{Undefined instruction} exception where the only subclasses of \texttt{Usage Fault} which could be observed. Both of these exceptions happen in the case of an invalid opcode. A possible explanation would be that a fault has been injected during the \textit{fetch} or \textit{decode} pipeline phases.

\subsubsection{Memory address}
In the initial state before the target instruction, \texttt{r0} points to a memory address in SRAM. The value of \texttt{r0} has been observed at this address instead of the expected value. For this particular case, instruction replacement simulation showed that the only possible instruction replacement is \texttt{STR r0,[r0,\#{}0]}, which stores the value of \texttt{r0} at the address pointed by \texttt{r0} without any offset. Moreover, the value \texttt{0x100} has also been observed for another configuration. It turns out that \texttt{0x100} is also the value in \texttt{r7}.


\subsubsection{Other faults}
We also obtained faults on the general-purpose register \texttt{r7} and the program counter \texttt{r15}. These faults can also be explained by at least one assembly instruction replacement.

\subsubsection{Summary}
Obviously, the previous paragraphs do not aim at providing a complete list of the possible faults. Because of the huge number of possible configurations for the injection parameters, computing fault occurrence percentages would not be relevant. Nevertheless, these paragraphs highlight the fact that very few fault patterns were observed. We never got any fault on \texttt{r1-r6} and \texttt{r8-r14}. In an informal way, faults on \texttt{r7} and \texttt{r15 (pc)} appeared much more often than faults on the memory address pointed by \texttt{r0}. Most of the 16-bit instructions can only manipulate the registers \texttt{r0} to \texttt{r7}. For example, a \texttt{MOVS r7, \#{}FF} operation is assembled into a 16-bit instruction, while a  \texttt{MOVS r8, \#{}FF} is assembled into a 32-bit instruction. In a 16-bit instruction, \texttt{r7} is encoded by a \texttt{111} binary sequence. The fact that registers \texttt{r0-r6} are encoded with a smaller number of \texttt{1} in their encoding slot might explain this higher fault occurrence rate on the \texttt{r7} register. Similarly, branch instructions have many \texttt{1} in their slot. As a conclusion for this set of experiments, every faulty result we observed has at least one instruction replacement which can explain it. The first intuition of a \textit{set at 1} fault model we saw for data fetches leads us to a more detailed analysis of the pipeline stages in section \ref{Section:Modele}.

\subsection{Faults on the data flow}
\label{Paragraphe:Fautes-DataFlow}

For this experiment, we targeted a single \texttt{LDR r4,[PC,\#{}44]} instruction. The inital value of \texttt{R4} was \texttt{0x0} and \texttt{PC$+$44} was a Flash memory address which contained \texttt{0x12345678} (this experiment used the same configuration as the one we had defined in \ref{Paragraphe:Impact-amplitude}). We obtained several faulty outputs such as \texttt{0xFE345678} or \texttt{0xFFF45678}. We consider that every fault which cannot be explained by an instruction replacement is a fault on the data flow. In this experiment, the target \texttt{LDR r4,[PC,\#{}44]} is a 16-bit instruction, followed by a 16-bit \texttt{NOP} instruction. The Thumb2 instruction set can only output a limited set of constants in a single data-processing instruction \cite{Thumb2}. Thus, some of the faulty output values we observe, such as \texttt{0xFFF45678}, could theoretically only be loaded with a single load from indirect register. Since the whole memory does not contain any \texttt{FFF4} pattern, a single load instruction could not explain this result. We performed an exhaustive search over the 16-bit and 32-bit instructions. No single instruction can lead to a result of \texttt{0xFFF45678}. However, fault injection might have had an impact on two 16-bit instructions. To handle this issue, we performed another experiment, in which the target instruction was a \texttt{LDR r8,[PC,\#{}44]}, with \texttt{0x12345678} stored at the address \texttt{PC$+$44}. Using \texttt{r8} instead of \texttt{r4} makes this instruction be assembled as a 32-bit instruction. Except the stack manipulation instructions, no 16-bit instruction can write a value into registers between \texttt{r8} and \texttt{r12}. For this new configuration, we were able to obtain several faulty values, such as \texttt{0xFFF45679} or \texttt{0xFFFC5679}. With an exhaustive simulation, we can now guarantee that no single instruction can lead to such a result. Since a part of the faulty value is similar to the expected one, we can assume this fault injection had an impact on the data flow.

\subsection{Analysis at a lower abstraction level}

Underpowering a circuit or overclocking it leads to the same kind of timing violation faults \cite{Zussa2012}, but knowing which among the clock tree or the power grid has been faulted is a tough task. To the best of our knowledge, recent research papers such as \cite{Poucheret2011} claimed that the coupling between the injection probe and the circuit lies mainly in the power distribution network. According to the experiments from the previous section, electromagnetic glitch fault injection seems to enable us to perform attacks whose effect is equivalent to voltage or clock glitches, with a local effect that enables us to target either the instruction bus or the data bus. The following section deeply studies the bus transfers and provides an explanation for the faults we observed at a register-transfer level.

\section{Register-transfer level fault model}
\label{Section:Modele}

\definecolor{green}{rgb}{0,0.4,0}
\begin{figure*}[!t]
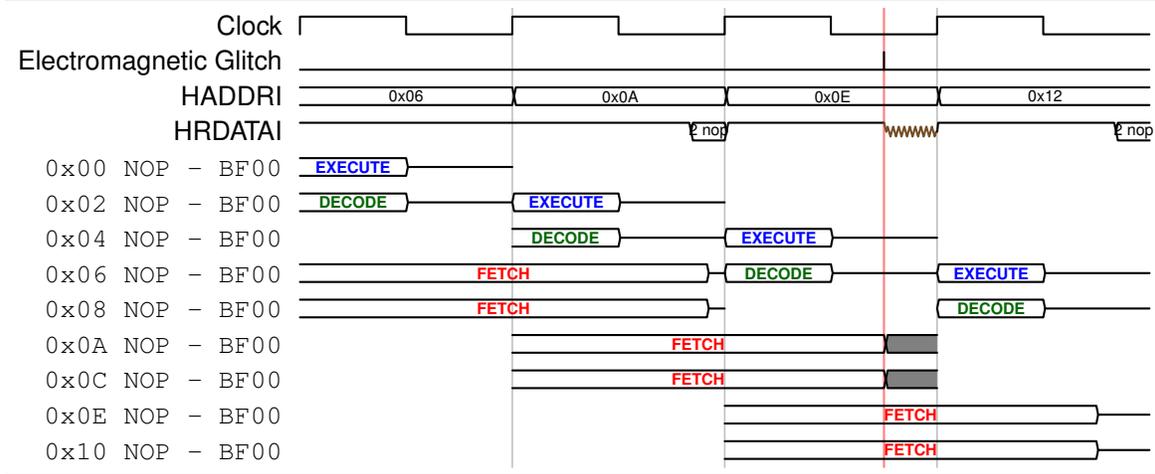

\centering
\begin{tikztimingtable}[scale=0.8]
  Clock  & [C]  8{6C} \\ 
  Electromagnetic Glitch & 33L G 15L \\
  HADDRI & {12D{0x06} 12D{0x0A} 12D{0x0E} 12D{0x12}}\\  
  HRDATAI & {12H 10H 2D{2 nop} 9H 3M 10H 2D{2 nop}} \\  
  {\tt 0x00 NOP - BF00} & {6D{\color{blue}\textbf{EXECUTE}} 6X[black]}\\
  {\tt 0x02 NOP - BF00} & {6D{\color{green}\textbf{DECODE}} 6X[black] 6D{\color{blue}\textbf{EXECUTE}} 6X[black]}\\
  {\tt 0x04 NOP - BF00} & {12S 6D{\color{green}\textbf{DECODE}} 6X[black] 6D{\color{blue}\textbf{EXECUTE}} 6X[black]}\\  
  {\tt 0x06 NOP - BF00} & {23D{\color{red}\textbf{FETCH}} 1X[black] 6D{\color{green}\textbf{DECODE}} 6X[black] 6D{\color{blue}\textbf{EXECUTE}} 6X[black]}\\  
  {\tt 0x08 NOP - BF00} & {23D{\color{red}\textbf{FETCH}} 1X[black] 12S 6D{\color{green}\textbf{DECODE}} 6X[black]}\\ 
  {\tt 0x0A NOP - BF00} & {12S 21D{\color{red}\textbf{FETCH}} 3U}\\  
  {\tt 0x0C NOP - BF00} & {12S 21D{\color{red}\textbf{FETCH}} 3U}\\ 
  {\tt 0x0E NOP - BF00} & {12S 12S 21D{\color{red}\textbf{FETCH}} 3X[black]}\\  
  {\tt 0x10 NOP - BF00} & {12S 12S 21D{\color{red}\textbf{FETCH}} 3X[black]}\\  
\extracode
  \tablerules
  \begin{pgfonlayer}{background}
    \begin{scope}[semitransparent,thick]
      \vertlines[red]{33}
    \end{scope}
    \begin{scope}[semitransparent,semithick]
          \vertlines[gray]{12,24,36}
    \end{scope}    
  \end{pgfonlayer}
\end{tikztimingtable}
\caption{Bus transfers on the AHB bus for instruction memory}
\label{Misc:Chronogramme-Instruction}
\end{figure*}

The results presented in section \ref{Section:Fautes-obtenues} lead us to the basics of a definition of a fault model at the assembly level. By using this fault injection technique, an attacker can inject faults in two ways: modify the instruction to be executed or modify a data value in the case of a load instruction.

Our experiments also highlight another trend: we only managed to inject faults on data and instruction transfers from the Flash memory. The Flash memory has a slower reponse time than the SRAM memory. The \textit{fetch} pipeline phase always requires a transfer from the instruction memory \cite{DefinitiveGuideARMCortexM3}. The operand fetch operation is performed during the \textit{decode} phase. For load instructions, the \textit{decode} phase requires a transfer from the data memory.

The microcontroller we use is based on a modified Harvard architecture, with separate buses for instruction and data. The buses are 32-bit wide and use the AMBA AHB-Lite structure \cite{AHB}. Since electromagnetic glitch injection creates timing faults \cite{Dehbaoui2012}, we propose an explanation of the experimental faults we obtained based on a bus transfer analysis.

\begin{figure*}[!t]
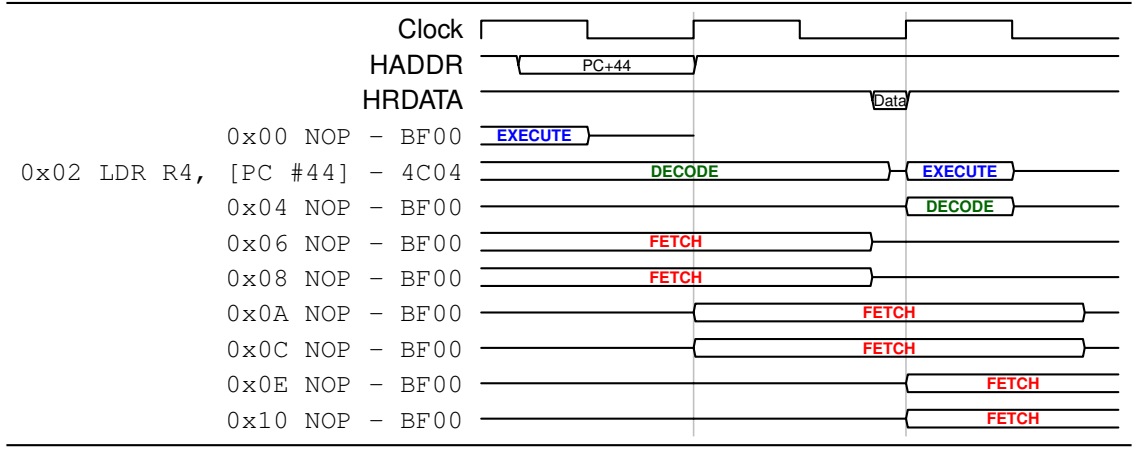
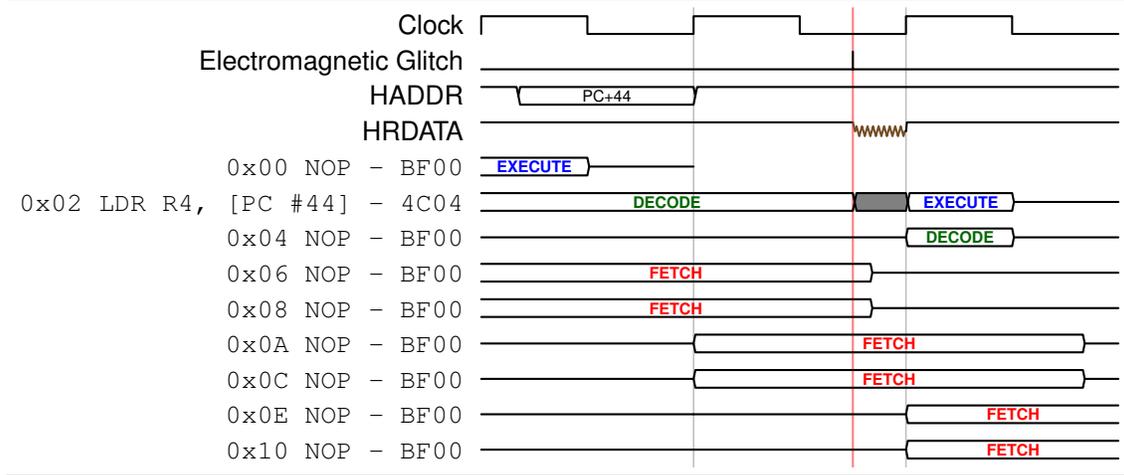

\centering
\begin{subfigure}[b]{\textwidth}
\centering
\begin{tikztimingtable}[scale=0.8]
  Clock  & [C]  6{6C} \\ 
  HADDR & {2H 10D{PC+44} 24H}\\  
  HRDATA & {12H 10H N(Data) 2D{Data} 12H}\\
  {\tt 0x00 NOP - BF00} & {6D{\color{blue}\textbf{EXECUTE}} 6X[black]}\\  
  {\tt 0x02 LDR R4, [PC \#44] - 4C04} & {23D{\color{green}\textbf{DECODE}} 1X[black] {6D{\color{blue}\textbf{EXECUTE}} 6X[black]}}\\
  {\tt 0x04 NOP - BF00} & {24X[black] 6D{\color{green}\textbf{DECODE}} 6X[black]}\\
  {\tt 0x06 NOP - BF00} & {22D{\color{red}\textbf{FETCH}} 2X[black] 12X[black]}\\  
  {\tt 0x08 NOP - BF00} & {22D{\color{red}\textbf{FETCH}} 2X[black] 12X[black]}\\ 
  {\tt 0x0A NOP - BF00} & {12X[black] 22D{\color{red}\textbf{FETCH}} 2X[black]}\\  
  {\tt 0x0C NOP - BF00} & {12X[black] 22D{\color{red}\textbf{FETCH}} 2X[black]}\\ 
  {\tt 0x0E NOP - BF00} & {12X[black] 12X[black] 12D{\color{red}\textbf{FETCH}}}\\  
  {\tt 0x10 NOP - BF00} & {12X[black] 12X[black] 12D{\color{red}\textbf{FETCH}}}\\  
\extracode
  \tablerules
  \begin{pgfonlayer}{background}

    \begin{scope}[semitransparent,semithick]
          \vertlines[gray]{12,24}
    \end{scope}    
  \end{pgfonlayer}
\end{tikztimingtable}
\caption{Without electromagnetic perturbation}
\label{Misc:Chronogramme-Data-SansEM}
\end{subfigure}

\begin{subfigure}[b]{\textwidth}
\centering
\begin{tikztimingtable}[scale=0.8]
  Clock  & [C]  6{6C} \\
  Electromagnetic Glitch & 21L G N(Glitch) 15L\\
  HADDR & {2H 10D{PC+44} 24H}\\  
  HRDATA & {12H 9H 3M 12H}\\
  {\tt 0x00 NOP - BF00} & {6D{\color{blue}\textbf{EXECUTE}} 6X[black]}\\  
  {\tt 0x02 LDR R4, [PC \#44] - 4C04} & {21D{\color{green}\textbf{DECODE}} 3U {6D{\color{blue}\textbf{EXECUTE}} 6X[black]}}\\
  {\tt 0x04 NOP - BF00} & {24X[black] 6D{\color{green}\textbf{DECODE}} 6X[black]}\\
  {\tt 0x06 NOP - BF00} & {22D{\color{red}\textbf{FETCH}} 2X[black] 12X[black]}\\  
  {\tt 0x08 NOP - BF00} & {22D{\color{red}\textbf{FETCH}} 2X[black] 12X[black]}\\ 
  {\tt 0x0A NOP - BF00} & {12X[black] 22D{\color{red}\textbf{FETCH}} 2X[black]}\\  
  {\tt 0x0C NOP - BF00} & {12X[black] 22D{\color{red}\textbf{FETCH}} 2X[black]}\\ 
  {\tt 0x0E NOP - BF00} & {12X[black] 12X[black] 12D{\color{red}\textbf{FETCH}}}\\  
  {\tt 0x10 NOP - BF00} & {12X[black] 12X[black] 12D{\color{red}\textbf{FETCH}}}\\  
\extracode
  \tablerules
  \begin{pgfonlayer}{background}
    \begin{scope}[semitransparent,thick]
      \vertlines[red]{21}
    \end{scope}
    \begin{scope}[semitransparent,semithick]
          \vertlines[gray]{12,24}
    \end{scope}    
  \end{pgfonlayer}
\end{tikztimingtable}
\caption{With an electromagnetic glitch fault injection}
\label{Misc:Chronogramme-Data-AvecEM}

\end{subfigure}
\caption{Bus transfers on the AHB bus for data memory}
\label{Misc:Chronogramme-Data}
\end{figure*}

\subsection{Instruction fetches}
Fetching a piece of data or an instruction from the memory (either SRAM or Flash) requires at least two clock cycles. Fig \ref{Misc:Chronogramme-Instruction} shows a chronogram of the AHB bus tranfers when executing the target program. In this case, the target program is a \texttt{NOP} sled. Since instruction fetches are 32-bit wide, two 16-bit \texttt{NOP} instructions are fetched at each execution of the \textit{fetch} pipeline stage. In the case of instruction which do not require an operand fetch, the \textit{decode} and \textit{execute} pipeline stages require at most  half a clock cycle. The \textit{fetch} stage requires one clock cycle during which the instruction address is written on the HADDRI bus. It also requires an extra clock cycle in which 32 bits from the instruction memory are written on the HRDATAI bus \cite{AHB}. In the event of a transfer from the SRAM memory, the values are written on the HRDATAI bus at the beginning of this extra clock cycle. Since the Flash memory has a longer response time, this value is written on the bus at the end of this clock cycle for a Flash transfer. In this situation, since electromagnetic fault injection leads to timing faults \cite{Dehbaoui2012}, the critical path appears to be this HRDATAI bus transfer. 

\begin{table}[!h]
\centering
\caption{Binary encoding of \texttt{NOP} and \texttt{STR r0,[r0,\#{}0]}}

\begin{tabular}{|l|l|l|l|}
   \hline
   \textbf{Mnemonic} & \textbf{Inst.} & \textbf{Binary instruction} & \textbf{Hamming w.} \\ 
   \hline
	\texttt{NOP} & BF00 & 10111111 00000000 & 7 \\
   \hline
	\texttt{STR r0,[r0,\#{}0]} & 6000 & 01100000 00000000 & 2 \\
   \hline
\end{tabular}
\label{Tableau:Opcodes}

\end{table}

We now consider the result presented in \ref{Paragraphe:Fautes-ProgramFlow}, in which a \texttt{NOP} is replaced by a \texttt{STR r0,[r0,\#{}0]}. The binary encodings for the \texttt{NOP} and \texttt{STR r0,[r0,\#{}0]} instructions are presented in Table \ref{Tableau:Opcodes}. As seen for an attack which targets the value loaded by a load instruction, it seems that the higher the stress we apply, the higher the fetched word's Hamming weight is. However, the situation seems different for instruction fetches. The fault models seems more complex than the \textit{set at 1} model we had seen for data fetches. 

The bus precharge values are not specified in the AHB bus intellectual property. They are chosen by the circuit's manufacturer. For the microcontroller we use, the HRDATAI bus does not seem to be precharged at 1, since a \texttt{NOP} instruction with a Hamming weight of 7 has been replaced by another instruction whose Hamming weight is 2. Moreover, since there must be some skew on this bus, some metastability phenomena (as presented in \ref{Paragraphe:Metastabilite}) also appear. Considering this single example, a possible precharge value would be 0, or the microcontroller might use a more complex precharge strategy. For the moment, we are not able to infer more details about a possible HRDATAI bus precharge. 

To sum up, in the case of a bus transfer from the Flash memory, the critical path that is faulted by an electromagnetic fault injection seems to be the HRDATAI bus transfer. Thus, this fault injection can target any instruction fetch from the Flash memory, which potentially makes this attack scenario very harmful.

\subsection{Data fetches}

The situation for data fetches is very similar to the one we describe for instruction fetches. Since electromagnetic fault injection has a local effect, it is possible to find a probe position where we only inject faults on the data bus and do not reach the instruction bus. For this attack scenario, the critical path seems to be the HRDATA bus transfer. Fig. \ref{Misc:Chronogramme-Data-SansEM} shows data bus transfers in the case of a single \texttt{LDR} instruction (similar to the previous experiments) without fault injection. Fig. \ref{Misc:Chronogramme-Data-AvecEM} shows the same bus transfers in the case of a fault injection.

Metastability phenomena also appear for this fault injection, but the global trend corresponds to a \textit{set at 1} fault model. This value depends on the microcontroller's bus precharge strategy, which is specific to each implementation. This trend enables us to define a more precise fault model for data fetches, in which the attacker can bring the loaded value closer to the value of the bus precharge.

\section{Related works}
\label{Section:Related works}
This section outlines some research papers that are related to the study we presented in this paper. These papers are grouped into three subcategories: electromagnetic fault injection techniques, fault models on microcontrollers and proposed contermeasures against a given fault model.

\subsubsection{Electromagnetic fault injection}
In \cite{Dehbaoui2012}, Dehbaoui \textit{et al.} do a practical fault injection on a software implementation of the AES algorithm by using electromagnetic glitches. In \cite{Carlier2012}, Carlier performs a study of the effects of electromagnetic fault injection on two microcontrollers at an electric level. His work mostly explains the influence of several parameters related to the coil. He also studies the influence of the injection time. However, his study does not focus on the faults that were produced.

\subsubsection{Fault models on microcontrollers}
In \cite{Barenghi2009}, Barenghi \textit{et al.} study the effects on low-voltage fault attacks on an ARM9 microprocessor. They describe several effects on loads from memory or on instruction replacement. In \cite{Balasch2011}, Balasch \textit{et al.} present a black-box approach which is quite similar to the one proposed in this paper. They use clock glitches as a fault injection mean and perform their experiment on a 8-bit microcontroller. We also use the same kind of in-depth analysis. However, their study is performed on a very different architecture with a different bus precharge configuration. We also automated the instruction replacement search by performing an exhaustive instruction replacement simulation over the instruction set. In \cite{Spruyt2012}, Spruyt proposes an approach whose aim is to define a generic method on how to build a fault model for microcontrollers. His situation is also quite similar to the one presented in this paper. Since having access to some internal data on a real microcontroller may be hard, he proposes a way to obtain information about the induced faults by analyzing the faulty outputs of different groups of instructions. Several articles have been published in which the authors assume an attacker can skip or replace an instruction by another one \cite{Schmidt2008} \cite{Schmidt2009}. For example, in \cite{Berzati2009}, Berzati \textit{et al.} suppose an attacker can replace an addition instruction by an exclusive or instruction. 

\subsubsection{Countermeasures}
Several countermeasures schemes have been defined to protect embedded processor architectures against specific fault models. All those countermeasure schemes might be reinforced by studies similar to the one presented in this paper, which could provide a more precise knowledge about the fault model. At a hardware level, \cite{Nguyen2011} proposes to use integrity checks to ensure that no instruction replacement took place. Nevertheless, many countermeasures to protect assembly code without modifying the microcontroller's architecture have been defined. In \cite{Barenghi2010}, Barenghi \textit{et al.} propose three countermeasure schemes based on instruction duplication, instruction triplication and parity checking. Their countermeasures enable different levels of fault detection and correction against instruction skips or some instruction modifications. In \cite{Medwed2008}, Medwed \textit{et al.} propose a generic approach based on the use of specific algebraic structures named \textit{AN+B} codes. Their approach enables to protect both the control and data flow. An application to an AES implementation has also been detailed in \cite{Medwed2010}. 

\section{Conclusion}
We have presented a detailed study about the effects of an electromagnetic glitch fault injection on a state-of-the-art microcontroller. However, working with a real microcontroller in a black-box approach creates several constraints when trying to build a practical experimental process. Because of the lack of details about the microcontroller's design and architecture, we have proposed this top-down approach which aims at building a suitable lower-level explanation for the faults we observed at an assembly level. Moreover, we also lack information about the bus precharge strategy on the microcontroller we use. Future experiments will try to use more advanced debug techniques in order to get more accurate information about the executed instructions.

Finally, we do not claim this register-transfer level hypothesis is the only reason why faults appear at an assembly level. This paper aims at providing a first understanding of the faults an electromagnetic glitch fault injection can induce on an embedded program. And the lower-level model we propose could explain all the previous experimental results we obtained. Furthermore, this fault model looks very similar to the ones which can be found in previous works about clock or voltage glitches. Hence, electromagnetic glitches seem to induce timing constraints violations on the bus transfers from the Flash memory. Thus, on a standard circuit, electromagnetic fault injection could enable an attacker to bypass some countermeasures against traditional timing fault injection means such as clock or voltage glitches. 

These experiments confirm the fact that an attacker could change an instruction into another one and change the value of a piece of data loaded from the Flash memory. But they also provide a more accurate fault model, in which some instructions or registers seem to be more vulnerable than others. On this architecture, faults on the data flow lead to an increased Hamming weight on the loaded piece of data. This behaviour highly depends on the microcontroller's bus precharge strategy. These observations can lead to the definition of an assembly-level fault model, and enable to build more specific and accurate countermeasures. These ideas will be studied more precisely in future works.


\balance
\bibliographystyle{IEEEtran}
\bibliography{IEEEabrv,library}

\end{document}